# Nimrod/G: An Architecture for a Resource Management and Scheduling System in a Global Computational Grid

Rajkumar Buyya†, David Abramson†, and Jonathan Giddy‡

| | |
|---|---|
| † School of Computer Science and Software Engineering<br>Monash University, Caulfield Campus<br>Melbourne, Australia | ‡ CRC for Enterprise Distributed Systems Technology<br>General Purpose South Building,<br>University of Queensland, St. Lucia<br>Brisbane, Australia |

Email: {rajkumar, davida, jon}@csse.monash.edu.au

**Abstract** - *The availability of powerful microprocessors and high-speed networks as commodity components has enabled high performance computing on distributed systems (wide-area cluster computing). In this environment, as the resources are usually distributed geographically at various levels (department, enterprise, or worldwide) there is a great challenge in integrating, coordinating and presenting them as a single resource to the user; thus forming a computational grid. Another challenge comes from the distributed ownership of resources with each resource having its own access policy, cost, and mechanism.*

*The proposed Nimrod/G grid-enabled resource management and scheduling system builds on our earlier work on Nimrod and follows a modular and component-based architecture enabling extensibility, portability, ease of development, and interoperability of independently developed components. It uses the Globus toolkit services and can be easily extended to operate with any other emerging grid middleware services. It focuses on the management and scheduling of computations over dynamic resources scattered geographically across the Internet at department, enterprise, or global level with particular emphasis on developing scheduling schemes based on the concept of computational economy for a real test bed, namely, the Globus testbed (GUSTO).*

## 1. Introduction

The popularity of the Internet and the availability of powerful computers and high-speed networks as low-cost commodity components are changing the way we do computing. This technology opportunity leads to the possibility of using networks of computers as a single, unified computing resource, popularly called cluster computing [7]. Clusters appear in various forms: high-performance clusters, high-availability clusters, dedicated clusters, non-dedicated clusters, and so on.

It is possible to cluster/couple wide varieties of geographically distributed resources such as computers including supercomputers, storage systems, data sources, and special class of devices and use them as a single unified resource, thus forming what is popularly known as "computational grids". The computational grid is analogous to the electric power grid, which provides pervasive access to power, and its vision is to offer dependable, consistent, pervasive and inexpensive access to high-end resources [18] irrespective of their physical location and the location of access points. This infrastructure is expected to have a transforming effect on society similar to that of the electric power grid and trigger the emergence of new classes of applications. Supercomputing is one of the driving applications for grid computing. Other applications include on-demand or just-in time, data intensive, collaborative, and high-throughput computing [23].

A large-scale parameter study of a simulation is well suited to high-throughput computing [4]. It involves the execution of a large number of tasks (task farms) over a range of parameters. There are many scientific and engineering applications that are used by domain/field experts to study the system behavior and they need several hours and days of compute time in order to perform system simulation to arrive at decision. Such experiments are usually conducted on expensive proprietary supercomputers costing millions of dollars to purchase and additional high maintenance cost. The cost of computing can be drastically reduced by using existing resources (such as PCs, workstations, SMPs, and clusters) scattered across the department, enterprise, or offices across the globe.

In the future, computational grids are expected to popularise a model of computational economy and create a market for those who are interested in offering computational resource rental services. In such an environment, whenever an application needs additional computational resources to run, the user can hire or rent resources on the fly and pay for what he uses. The resource price may vary from time to time and from one user to another user. At runtime, the user can even enter into bidding and negotiate for the best possible resources for low-cost access from computational service providers. In order to get the best value for money, the user can reserve the resources in advance.

The use of a computational economy can be made simpler by providing a layer that allows the user to select a "deadline", the period within which an application execution must be completed, and a "price", the amount that the user is willing to pay for the completion of the application. This layer is then responsible for agglomerating individual resources to satisfy these constraints (user requirements) using resource reservation and bidding methods.

In order to address the complexities associated with parametric computing on clusters of distributed systems, we have devised a system called Nimrod [1][2][3][4]. Nimrod provides a simple *declarative* parametric modeling language for expressing a parametric experiment. The domain experts can easily create a *plan* for a parametric computing (task farming) and use the Nimrod runtime system to submit, run, and collect the results from multiple computers (cluster nodes). Nimrod has been used to run applications ranging from bio-informatics and operations research to the simulation of business processes [12][21]. A reengineered version of Nimrod, called Clustor, has been commercialized by Active Tools [5].

The Nimrod system has been used successfully with a static set of computational resources, but is unsuitable as implemented in the large-scale dynamic context of computational grids, where resources are scattered across several administrative domains, each with their own user policies, employing their own queuing system, varying access cost and computational power. These shortcomings are addressed by our new system called Nimrod/G that uses the Globus [17] middleware services for dynamic resource discovery and dispatching jobs over computational grids.

The preliminary monolithic version of Nimrod/G has been discussed in [4]. In this paper we mainly discuss the architecture of a new highly modularized, portable, and extensible version of Nimrod/G. It takes advantage of features supported in the latest version (v1.1) of Globus [20] such as automatic discovery of allowed resources. Furthermore, we introduce the concept of computational economy as part of the Nimrod/G scheduler. The architecture is extensible enough to use any other grid-middleware services such as NetSolve [8]. The rest of the paper focuses on Nimrod/G architecture and its interactions with grid components, scheduling, computational economy, and related work.

## 2. System Architecture

The architecture of Nimrod/G is shown in Figure 1 and its key components are:
- Client or User Station
- Parametric Engine
- Scheduler
- Dispatcher
- Job-Wrapper

The interaction between the above components and grid resources is shown in Figure 2.

**Client or User Station**

This component acts as a user-interface for controlling and supervising an experiment under consideration. The user can vary parameters related to time and cost that influence the direction the scheduler takes while selecting resources. It also serves as a monitoring console and lists status of all jobs, which a user can view and control. Another feature of the Nimrod/G client is that it is possible to run multiple instances of the same client at different locations. That means the experiment can be started on one machine, monitored on another machine by the same or different user, and the experiment can be controlled from yet another location. We have used this feature to monitor and control an experiment from Monash University in Australia and Argonne National Laboratory in the USA simultaneously. It is also possible to have alternative clients, such as a purely text-based client, or another application (Active Sheets [16], an extended Microsoft Excel spreadsheet that submits cell functions for execution on the computational grid).

**Parametric Engine**

The parametric engine acts as a persistent job control agent and is the central component from where the whole experiment is managed and maintained. It is responsible for parameterization of the experiment and the actual creation of jobs, maintenance of job status, interacting with clients, schedule advisor, and dispatcher. The parametric engine takes the experiment plan as input described by using our *declarative* parametric modeling language (the plan can also be created using the Clustor GUI [13]) and manages the experiment under the direction of schedule advisor. It then informs the dispatcher to map an application task to the selected resource.

The parametric engine maintains the state of the whole experiment and ensures that the state is recorded in persistent storage. This allows the experiment to be restarted if the node running Nimrod goes down. The parametric engine exposes the Clustor network interface [14] to the other components and allows new components to be "plugged in" to the central engine.

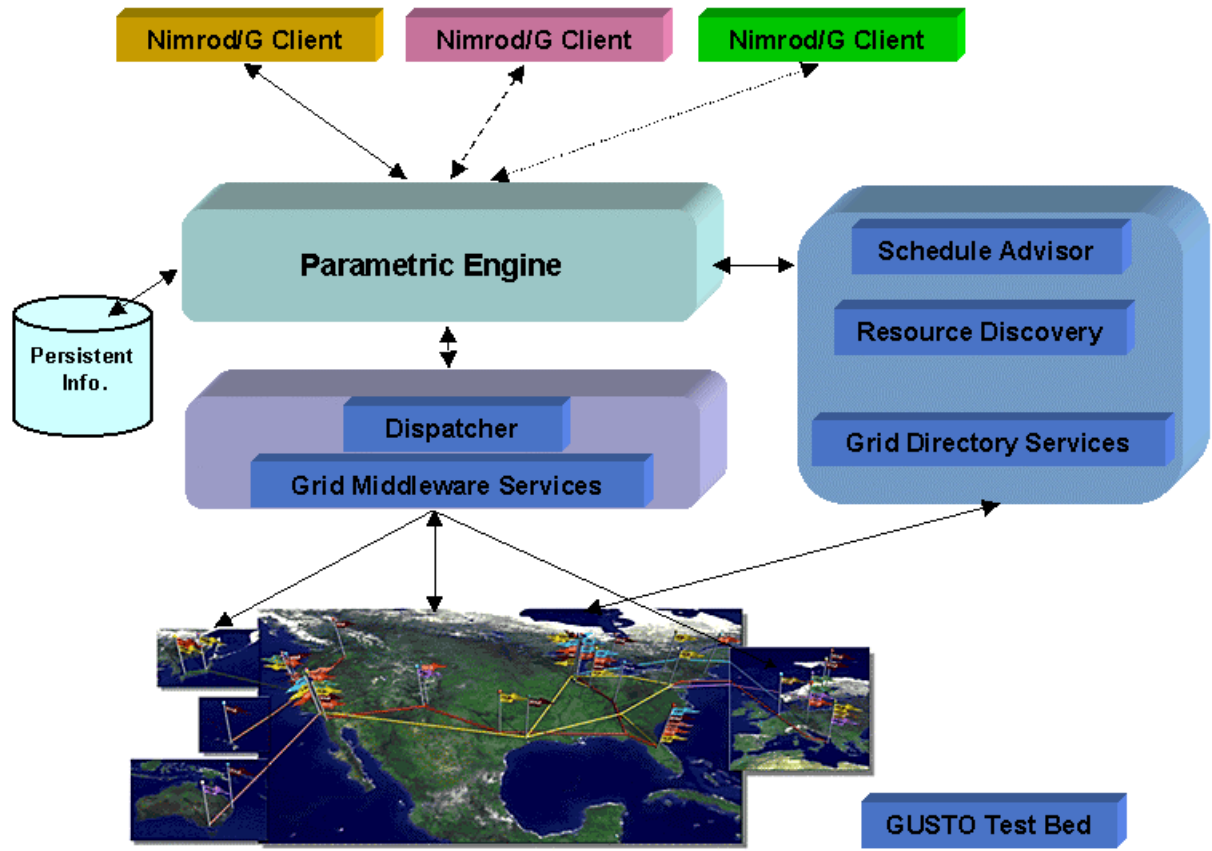

**Figure 1. The Architecture of Nimrod/G**

**Scheduler**

The scheduler is responsible for resource discovery, resource selection, and job assignment. The resource discovery algorithm interacts with a grid-information service directory (the MDS in Globus), identifies the list of authorized machines, and keeps track of resource status information. The resource selection algorithm is responsible for selecting those resources that meet the deadline and minimize the cost of computation. For further discussion on scheduling and computational economy, see Section 3.

**Dispatcher**

The dispatcher primarily initiates the execution of a task on the selected resource as per the scheduler's instruction. It periodically updates the status of task execution to the parametric-engine. In the current implementation, the dispatcher starts a remote component, known as job-wrapper. The job wrapper interprets a simple script containing instructions for file transfer and execution subtasks.

**Job Wrapper**

The job-wrapper is responsible for staging of application tasks and data; starting execution of the task on the assigned resource and sending results back to the parametric engine via dispatcher. It basically mediates between parametric engine and the actual machine on which the task runs.

## 3. Scheduling and Computational Economy

The integration of computational economy as part of a scheduling system greatly influences the way computational resources are selected to meet the user requirements. It can be handled in two ways. First, systems like Nimrod/G can work on the user's behalf and try to complete the assigned work within a given deadline and cost. The deadline represents a time by which the user requires the result, and is often imposed by external factors like production schedules or research deadlines. Second, the user can enter into a contract with the system and pose requests such as "this is what I am willing to pay if you can complete the job within the deadline". In the second case, the user can negotiate for resources in the grid and find out if the job can be performed. The system can employ resource reservation or a trading technique in order to identify suitable resources. Then the user can either proceed or renegotiate either by changing the deadline and/or the cost. The advantage of this approach is that the user knows before the experiment is started whether the system can deliver the results and what the cost will be.

To some extent our earlier prototype of Nimrod/G has followed the first method and was able to select resources based on artificial costs [4]. This system tries to find sufficient resources to meet the user's deadline, and adapts the list of machines it is using depending on competition for them. However, the cost changes as other competing experiments are put on the grid. The implementation of the second method of computational economy as part of our Nimrod/G system is rather complex and it needs grid middleware services for resource reservation, broker services for negotiating cost, and the underlying system having management and accounting infrastructure in place.

The scheduling system can use various kinds of parameters in order to arrive at a scheduling policy to optimally complete an application execution. The parameters to be considered include,

- Resource Architecture and Configuration
- Resource Capability (clock speed, memory size)
- Resource State (such as CPU load, memory available, disk storage free)
- Resource Requirements of an Application
- Access Speed (such as disk access speed)
- Free or Available Nodes
- Priority (that the user has)
- Queue Type and Length
- Network Bandwidth, Load, and Latency (if jobs need to communicate)
- Reliability of Resource and Connection
- User Preference
- Application Deadline
- User Capacity/Willingness to Pay for Resource Usage
- Resource Cost (in terms of dollars that the user need to pay to the resource owner)
- Resource Cost Variation in terms of Time-scale (like high @ daytime and low @ night)
- Historical Information, including Job Consumption Rate

The important parameters of computational economy that can influence the way resource scheduling is done are:

- Resource Cost (set by its owner)
- Price (that the user is willing to pay)
- Deadline (the period by which an application execution need to completed)

The scheduler can use all sorts of information gathered by a resource discoverer and also negotiate with resource owners to get the best "value for money". The resource that offers the best price and meets resource requirements can eventually be selected. This can be achieved by resource reservation and bidding. If the user deadline is relaxed, the chances of obtaining low-cost access to resources are high. The cost of resources can vary dynamically from time to time and the resource owner will have the full control over deciding access cost. Further, the cost can vary from one user to another. The scheduler can even solicit bids or tenders from computational resource providers in an open market, and select the feasible service-provider and use. It is real challenge for the resource sellers to decide costing in order to make profit and attract more customers.

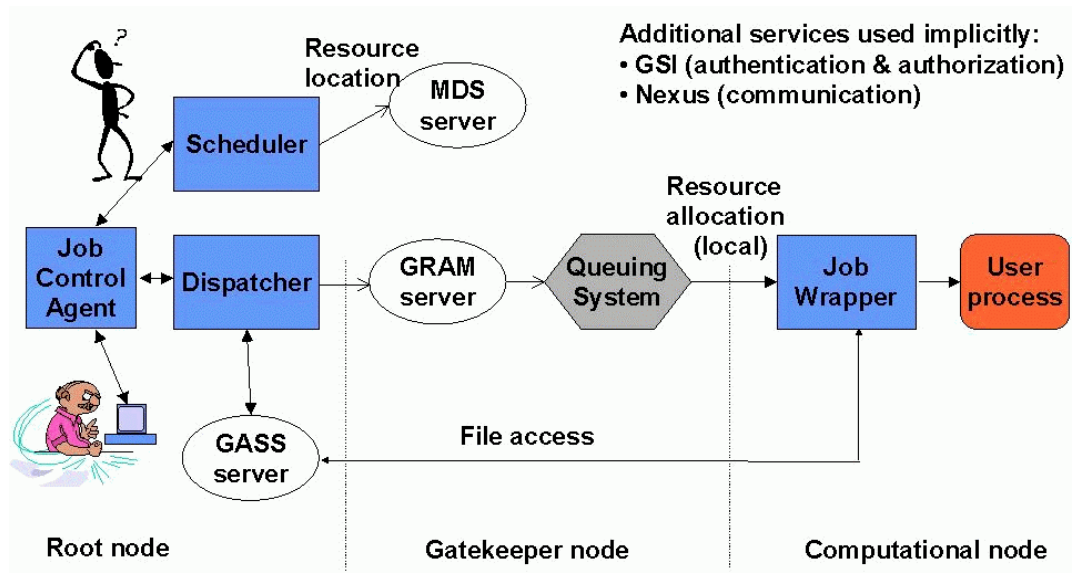

**Figure 2: Nimrod/G and Globus Components Interactions**

## 4. Implementation

The interaction between various components of Nimrod/G and grid resources (see Figure 2) has been discussed in the above sections and from this it is clear that they need to use dedicated protocols for communication. Nimrod/G components use TCP/IP sockets for exchanging commands and information between them. The implementation of the latest version of Nimrod/G follows the Clustor network protocols [14] as much as possible. In order to avoid the need for every user to understand the low-level protocols, we are developing a library of higher level APIs that can be accessed by any other extension components inter-operating with the Nimrod/G job control agent. A user could build an alternative scheduler by using these APIs.

The earlier version of Nimrod/G [4] was developed using Globus Toolkit (version 1.0). The components of Globus used in its implementation are: GRAM (Globus Resource Allocation Manager), MDS (Metacomputing Directory Service), GSI (Globus Security Infrastructure), and GASS (Global Access to Secondary Storage). The current version also uses these services along with the new features (such as Grid Directory Information Services) supported by the latest version of Globus Toolkit (version 1.1) [20].

In the future, the Globus toolkit is expected to support resource reservation services [19]. Our scheduler will use those services in order to support the market-based computational-economy model discussed earlier. However, currently we plan to build a simulated model for investigation purposes and build a real model when middleware services are made available.

In many dedicated clusters of computers (such as Beowulf-class Linux clusters) it is common for only the master node to be able to communicate to the external world (Internet) and all other nodes are interconnected using private (high-speed) networks. Accordingly, these private nodes of the cluster can only be accessed by master-node. In order to address this problem, we have developed a proxy server in order to integrate closed cluster nodes as part of computation grids. The proxy deployed on the cluster master node acts as a mediator between external Nimrod components and cluster private-nodes for accessing storage. When a client, running on a cluster private nodes, makes I/O call for accessing data available on external system, the proxy uses Globus GASS services to fetch or stage the required data.

## 5. Evaluation

The purpose of this paper is to present a new component based architectural design for Nimrod/G that will allow it to be extended and implemented using different middleware services. The new architecture will also support experimentation with a computational economy in a way that has not been possible to date. In order to illustrate the potential for such a system, we show the results produced using the previous monolithic version of Nimrod/G [4] using a real science case study—simulation of an Ionization Chamber Calibration on the GUSTO testbed resources. In this experiment, we ran the code across different design parameters and specified different real time deadlines. For the sake of completeness, the results of trials conducted during April/May 1999 are shown in Figure 3. There were about 70 machines available to us during the trial. We expect to repeat this trial using the new architecture in the near future, and then experimentation will begin with a new computational economy as discussed in Section 3.

Figure 3 shows the effect of varying the deadline for the experiment on the number of processors used. Not surprisingly, as the deadline is tightened, the scheduler needs to find more resources until the deadline can be met. The GUSTO test-bed resources selected change from one deadline to another and also from time to time due to the variation of availability/status of resources. When the deadline is tight, the scheduler selects a large number of resources (even though they are likely to be expensive) in order to complete the experiment within the deadline. In the above case, the scheduler has selected resources to keep the cost of experiment as low as possible, yet meeting the deadline. This clearly demonstrates the ability and the scalability of Nimrod/G to schedule tasks according to time and cost constraints over grid resources. The new architecture will allow more varied experimentation.

## 6. Related Work

A number of projects are investigating scheduling on computational grids. They include AppLeS [6], NetSolve [8], and DISCWorld [15], but these do not employ the concept of computational economy in scheduling. REXEC [10] supports the concept of computational economy, but it is limited to department/campus-wide network of workstations.

The AppLeS (Application-Level Scheduling) builds agents for each application (case-by-case) responsible for offering a scheduling mechanism [6]. It uses the NWS (Network Weather Service)[22] to monitor the varying loads on resources/networks to select viable resource configurations. Whereas, Nimrod/G offers a tool level solution that applies to all applications and the users are not required to build scheduling agents for each of their applications as in AppLeS.

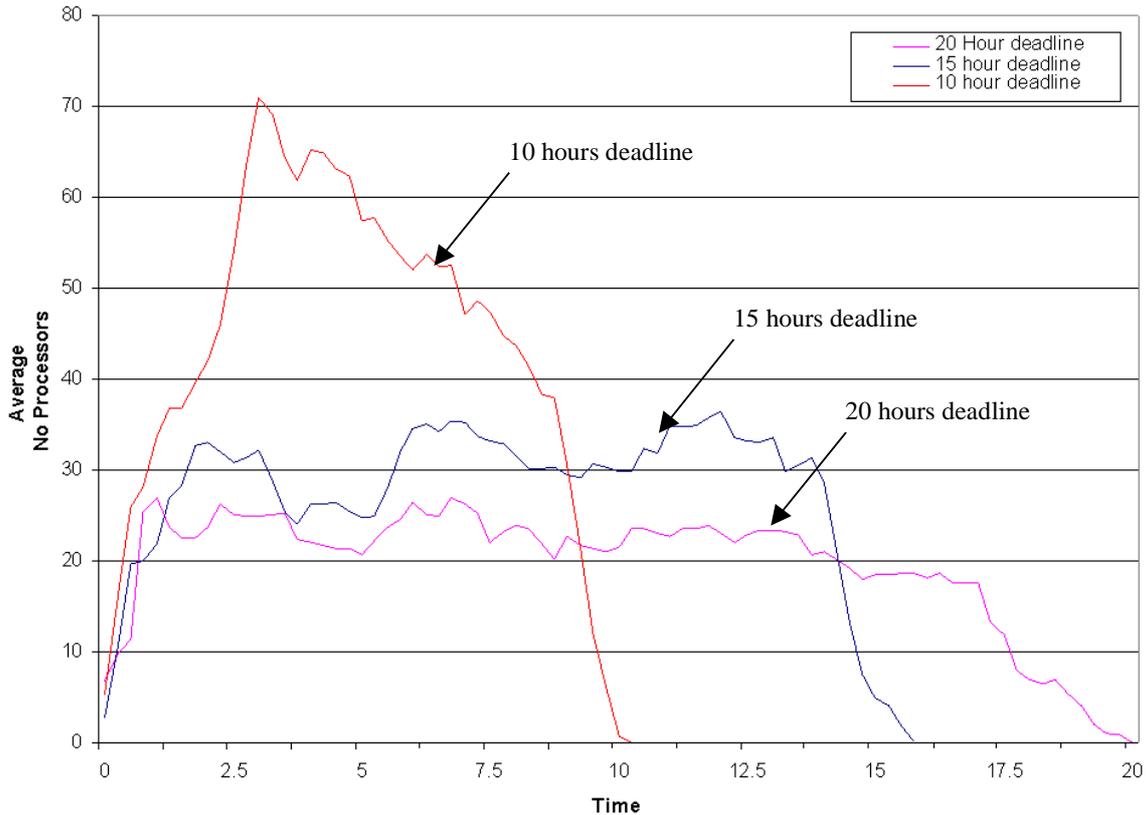

**Figure 3. GUSTO resources usage for 10, 15, and 20 hours of deadline. [4]**

NetSolve is a client-agent-server system, which enables the user to solve complex scientific problems remotely [8]. The NetSolve agent does the scheduling by searching for those resources that offer the best performance in a network. The applications need to be built using one of the APIs provided by NetSolve to perform RPC-like computations. NetSolve also provides an API for creating *task farms* [9] that means, the framing applications needs to be developed on a case-by-case basis using NetSolve APIs. In case of Nimrod, non computer-scientists can easily use the GUI to create task farms without modifying the application. However, it is interesting to note that the NetSolve middleware services can be used to build a system like Nimrod/G parametric engine. But the concept of computational economy cannot be supported unless NetSolve offers the ability to build scheduling agents or allows the user to plug-in their own scheduling policies.

DISCWorld (Distributed Information Systems Control World) is a service-oriented metacomputing environment, based on the client-server-server model [15]. Remote users can login to this environment over the Internet and request access to data, and also invoke services or operations on the available data. DISCWorld aims for remote information access whereas, Nimrod/G focuses on providing an easy and transparent mechanism for accessing computational resources.

Another related tool that uses the concept of computational economy is REXEC, a remote execution environment [10] for a campus-wide network of workstations, which is part of the Berkeley Millennium project. At the command line, the user can specify the maximum rate (credits per minute) that he is willing to pay for CPU time. The REXEC client selects a node that fits the user requirements and executes the application on it. The REXEC provides an extended shell for remote execution of applications on clusters. That is, it offers a generic user interface for computational economy on clusters, whereas Nimrod/G aims at offering a comprehensive and specialized environment for parametric computing in computational grids. Another difference is that REXEC executes jobs on the node computer directly, whereas Nimrod/G has the capability to submit jobs to queues on a remote system that, in turn, manages compute resources.

## 7. Conclusions and Future Work

The evolution of Nimrod from scheduling for a local computing environment to scheduling for the global computational grid has been discussed. It particular, we focused on the Nimrod/G architecture for resource management and scheduling on a computational grid. The various parameters that influence the scheduling on computational grids with a computational economy are presented. Some preliminary results from an earlier experiment are presented to demonstrate the scalability and the ability of Nimrod/G in making good scheduling decisions. Finally, we show how Nimrod/G relates to other projects.

The future work focuses on the use of economic theories in grid resource management and scheduling. The components that make up G̲rid A̲rchitecture for C̲omputational E̲conomy (GRACE) include global scheduler (broker), bid-manager, directory server, and bid-server working closely with grid middleware and fabrics. The GRACE infrastructure also offers generic interfaces (APIs) that the grid tools and applications programmers can use to develop software supporting the computational economy.

## Acknowledgments

The Nimrod project is supported by the Distributed Systems Technology Centre (DSTC) under the Australian Government CRC program. The award of the Australian Government International Postgraduate Research Scholarship (IPRS), the Monash University Graduate Scholarship (MGS), the DSTC Monash Scholarship, and the IEEE Computer Society Richard E Merwin Scholarship is acknowledged.

We thank Jack Dongarra (University of Tennessee, Knoxville), Francine Berman (University of California, San Diego), Toni Cortes (Universitat Politecnica de Catalunya, Barcelona), Hai Jin (University of Southern California, Los Angeles), Savithri S (Motorola India Electronics Ltd., Bangalore), and anonymous reviewers for their comments on the work.